\documentclass[lettersize,journal]{IEEEtran}

\usepackage{setspace}
\usepackage{amsmath,amsfonts}

\usepackage{algorithm}
\usepackage{algpseudocode}
\usepackage{array}
\usepackage[caption=false,font=normalsize,labelfont=sf,textfont=sf]{subfig}
\usepackage{textcomp}
\usepackage{stfloats}
\usepackage{url}
\usepackage{verbatim}
\usepackage{graphicx}
\usepackage{cite}
\usepackage{multirow}
\usepackage{booktabs}
\usepackage{amsmath,amssymb,amsfonts}
\usepackage{colortbl}
\usepackage{setspace}
\usepackage{authblk}
\begin{document}

\title{DP-TRAE: A Dual-Phase Merging Transferable Reversible Adversarial Example for Image Privacy Protection}







\author{
Xia Du, Jiajie Zhu, Jizhe Zhou$^*$, Chi-man Pun, ~\IEEEmembership{Senior Member,~IEEE}, Zheng Lin, Cong Wu,~\IEEEmembership{Member,~IEEE}, Zhe Chen,~\IEEEmembership{Member,~IEEE}, and Jun Luo,~\IEEEmembership{Fellow,~IEEE} 
        
\thanks{The part of this work has been published in ACM MM 2024.}
\thanks{Xia Du and Jiajie Zhu are with the School of Computer and Information Engineering, Xiamen University of Technology, Xiamen, 361000, China (email: duxia@xmut.edu.cn; cosmics36@163.com).}
\thanks{Jizhe Zhou is with the School of Computer Science, Engineering Research Center of Machine Learning and Industry Intelligence, Sichuan University, Chengdu, China, 610020, China (email: yb87409@um.edu.mo).}
\thanks{Chi-man Pun is with the Department of Computer and Information Science, Faculty of Science and Technology, University of Macau, Macau, 999078, China (email: cmpun@umac.mo). }
\thanks{Zheng Lin and Cong Wu are with the Department of Electrical and Electronic Engineering, University of Hong Kong, Pok Fu Lam,
Hong Kong SAR, China (e-mail: linzheng@eee.hku.hk; congwu@hku.hk).}
\thanks{Zhe Chen is with the Institute of Space Internet, Fudan University, Shanghai, China, and the School of Computer Science, Fudan University, Shanghai, China (e-mail: zhechen@fudan.edu.cn).}
\thanks{Jun Luo is with the School of Computer Science and Engineering, Nanyang
Technological University, Singapore (e-mail: junluo@ntu.edu.sg).}
\thanks{$^*$ denotes Corresponding author.}

\thanks{
Corresponding author: Jizhe Zhou (yb87409@um.edu.mo).
}
}

\markboth{Journal of \LaTeX\ Class Files,~Vol.~14, No.~8, August~2021}%
{Shell \MakeLowercase{\textit{et al.}}: A Sample Article Using IEEEtran.cls for IEEE Journals}


\maketitle
\begin{abstract}
In the field of digital security, Reversible Adversarial Examples (RAE) combine adversarial attacks with reversible data hiding techniques to effectively protect sensitive data and prevent unauthorized analysis by malicious Deep Neural Networks (DNNs). However, existing RAE techniques primarily focus on white-box attacks, lacking a comprehensive evaluation of their effectiveness in black-box scenarios. This limitation impedes their broader deployment in complex, dynamic environments. Furthermore, traditional black-box attacks are often characterized by poor transferability and high query costs, significantly limiting their practical applicability. To address these challenges, we propose the Dual-Phase Merging Transferable
Reversible Attack method, which generates highly transferable initial adversarial perturbations in a white-box model and employs a memory-augmented black-box strategy to effectively mislead target models. Experimental results demonstrate the superiority of our approach, achieving a 99.0\% attack success rate and 100\% recovery rate in black-box scenarios, highlighting its robustness in privacy protection. Moreover, we successfully implemented a black-box attack on a commercial model, further substantiating the potential of this approach for practical use.
\end{abstract}


\section{Introduction}
\IEEEPARstart{D}{eep} Neural Networks (DNNs) have initiated a technological revolution in various fields \cite{DNN1,DNN2,lin2024fedsn,fang2024automated,zhang2025lcfed,hu2024agentscodriver,fang2024ic3m,lin2025hierarchical,hu2024accelerating,yuan2023graph,yuan2025constructing,peng2025sigchord,lin2022channel,tang2024merit}, such as image recognition, natural language processing, and autonomous driving. Despite rapid progress in artificial intelligence across these domains, concerns about security and privacy have also increased \cite{liu2021seeing,curzon2021privacy,meng2022adversarial,acm2024machine}. Malicious actors often exploit unauthorized DNNs to analyze and steal users' private data to their advantage \cite{ni2023xporter,malicious1,ni2023exploiting,duan2025rethinking,ni2023uncovering}. For example, Cambridge Analytica used unauthorized personal data from Facebook users for targeted political advertising \cite{facebook}.

\begin{figure}[t]
  \centering
  \includegraphics[width=0.9\linewidth]{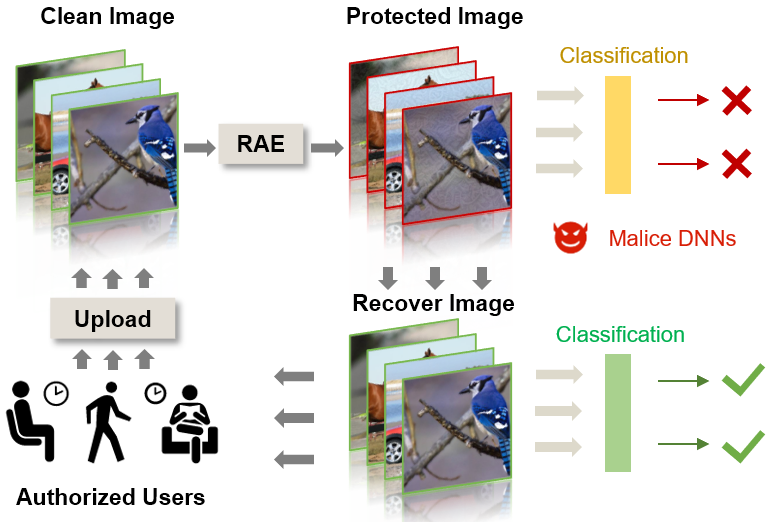}
  \caption{RAEs prevent malicious DNNs from stealing privacy data and can recover the image quality when necessary.}
  \label{figure1} 
\end{figure}

Recent research has exposed critical vulnerabilities in DNNs \cite{jia2022subnetwork,wang2023towards,naseer2021generating}, particularly their susceptibility to variations in input data quality and distribution \cite{ding2022interpreting}. In image recognition tasks, even minor pixel modifications can significantly mislead classification results \cite{naseer2019cross,wang2021feature,yuan2024itpatch,zhu2022toward}. Adversarial attacks exploit these weaknesses by introducing subtle perturbations that mislead DNNs incorrect predictions \cite{szegedy2013intriguing}. Recent studies have proposed using adversarial attacks to protect image privacy from malicious DNN analysis \cite{oh2017adversarial,liu2023diffprotect,xue2021socialguard}. However, the adversarial noise introduced directly is often irreversible, leading to degraded image quality and reduced data usability, especially for digital images. Therefore, there is an urgent need for a technique that can protect image privacy while preserving visual quality. Our research indicates that Reversible Adversarial Examples (RAEs) \cite{RAEliu2023unauthorized} offer significant potential for data protection: this approach ensures data security while enabling the adversarial perturbations to be reversed, restoring the original image. As illustrated in Figure \ref{figure1}, RAEs can generate controlled and reversible adversarial perturbations to effectively mislead unauthorized DNNs, thus safeguarding user privacy without compromising data usability.

The field of RAEs is still in its early stages of development. Liu \textit{et al.} first introduced the concept of reversible adversarial attacks by integrating reversible data hiding techniques with adversarial examples \cite{RAEliu2023unauthorized}. Xiong \textit{et al.} further expanded reversible adversarial attacks to black-box scenarios, employing ensemble model techniques to demonstrate the potential of RAEs across multiple models \cite{RAExiong2023black}. Although their approach exhibited strong transferability and misleading capability, it faced limitations in effectively attacking previously unexploited models. Meanwhile, Zhang \textit{et al.} \cite{RAEzhang2022self} proposed a method utilizing RGAN technology to replace reversible data hiding techniques, efficiently generating adversarial examples via an attacking encoder network and reversing them through a recovery decoder network. While this approach successfully restored the original images, it was less effective against unknown models.

Due to the specific nature of RAE, most current RAEs rely on the transferability of white-box attacks to achieve attacks. However, the effectiveness of this transferability is predominantly contingent upon the intensity of the applied perturbations. The RDH method imposes strict limitations on the magnitude of perturbations, leading to a design conflict that significantly reduces the success rate of attacks against unknown black-box models.

To overcome the potential challenges of existing RAE techniques, particularly the balance between implementing effective adversarial attacks and adhering to the strict perturbation limits of reversible data hiding \cite{yin2020reversible,qian2016separable} techniques, we propose the Dual-Phase Merging Transferable Reversible Adversarial Example (DP-TRAE), a novel reversible adversarial attack on black-box models. The preliminary version \cite{DP-RAE} of this work was presented at ACM MM 2024. DP-TRAE divides the entire attack into two phases: the Stepwise Adaptive  White-box Attack (SA-WA) and a Memory-Assisted Expansion Black-box Attack (MAE-BA). The motivation behind DP-TRAE is to combine the high transferability of white-box attacks with the targeted nature of black-box attacks, thus reducing the overall cost of the attack. Our intuition suggests that, compared to random perturbations, conducting a black-box attack on top of adversarial perturbations initialized by a white-box attack can mislead the target model more efficiently, similar to modifying an existing masterpiece with a clear direction in mind. Although white-box adversarial noise may not fully align with the gradient ascent direction of an unknown model, it provides a superior initial direction for the attack. Among them, The SA-WA method introduces additional perturbation guidance for gradient-sensitive regions, thereby enhancing the efficiency of misdirection. It adaptively adjusts the magnitude of perturbations to mitigate overfitting, ensuring more robust attack performance. MAE-BA method records the impact of each perturbation on the results, accumulating and utilizing historical data to select the most promising update points and enhance the perturbation intensity in neighboring regions. Furthermore, to address the conflict between perturbation and RDH storage capacity \cite{RAEliu2023unauthorized, RAExiong2023black, RAEyin2019reversible}, we regularize the perturbations and compress them using Huffman coding, effectively mitigating the trade-off between perturbation intensity and storage requirements. In particular, the main contributions of this work are as follows:

$\bullet $ We propose a novel Adaptive Transferable Reversible Adversarial Attack framework for black-box attacks, integrating multiple attack strategies to enhance both transferability and efficiency effectively. To the best of our knowledge, our approach is the first successful application of RAE attacks on commercial black-box models. 

$\bullet $ We propose the MAE-BA and SA-WA to accelerate the generation and compression of effective perturbations and employ Huffman coding to further compress the final perturbation information. These approaches tackle the key challenge of preserving the integrity of adversarial examples while guaranteeing their reversibility.

$\bullet $ Experimental results affirm the superiority of our attack framework, achieving a 99.0\% Attack Success Rate (ASR) for reversible adversarial examples on specific models, with a 100\% restoration rate for the recovered images. These outcomes validate the practical feasibility of our approach.

The rest of this paper is organized as follows. Section II describes the relevant background and technology. Section III provides the detailed design of the DP-TRAE. Section IV fully analyzes the experimental results and verifies the feasibility of the proposed DP-TRAE. Finally, the conclusions and future outlooks are presented in Section V.

\section{Related Work and Background}
In this section, we review the existing work on adversarial attacks and the related techniques used in the proposed methods.

\subsection{Adversarial Attack}
Adversarial attacks generate perturbations to mislead model decisions. Goodfellow \textit{et al.} \cite{FGSM} introduced the Fast Gradient Sign Method (FGSM), which exposed the vulnerability of deep learning models by adding small perturbations along the gradient direction. Despite FGSM's computational efficiency, its performance in complex scenarios is limited. To address these limitations, Kurakin \textit{et al.} \cite{IFGSM} proposed the iterative-FGSM (I-FGSM), enhancing attack effectiveness through multiple minor iterative updates. Dong \textit{et al.} \cite{MIFGSM} extended I-FGSM by incorporating a momentum term, resulting in the Momentum Iterative Method (MI), which stabilizes the update direction and significantly improves the transferability of adversarial examples across different models.

Attacks based on a single input often exhibit poor transferability due to overfitting to a specific model. To alleviate this issue, data augmentation techniques have been integrated into adversarial attacks to increase input diversity. Xie \textit{et al.} proposed the Diverse Input Method (DI) \cite{DIM}, which enhances adversarial attack effectiveness by applying random transformations (e.g., scaling and cropping). Dong \textit{et al.} \cite{TIM} further proposed the Translation-Invariant Method (TI), which uses convolution to smooth the gradient, expanding the perturbation's spatial extent and improving generalizability to unseen models.

In the black-box attack setting, Guo \textit{et al.} developed the Simple Black-box Attack (SimBA) \cite{Simba}, which modifies individual input dimensions to evaluate their impact on model output, generating compelling adversarial examples with minimal queries. This straightforward approach demonstrates the efficiency of black-box attacks without relying on gradient information. On the other hand, Maho \textit{et al.} proposed the SurFree attack optimizes the query path using geometric considerations \cite{Surfree}, eliminating the need for a substitute model. This approach substantially reduces query complexity while maintaining high attack success rates.

\subsection{Reversible Data Hiding}

Image steganography involves embedding secret information within images in an imperceptible way to the human eye. The Least Significant Bit (LSB) \cite{LSB} substitution is a popular and highly effective technique known for its simplicity and ease of implementation while maintaining high imperceptibility. Discrete Cosine Transform embeds data in the frequency domain, providing robustness against compression. Discrete Wavelet Transform offers better localization in both spatial and frequency domains \cite{DCT}. Grayscale invariant reversible steganography, which allows perfect image recovery after data extraction while maintaining robustness to grayscale variations, has also emerged as an important method \cite{Gray}. Recently, deep learning-based approaches have been introduced to further enhance the security and capacity of steganographic systems \cite{jing2021hinet, zhu2018hidden}. In a nutshell, it can be described as:
\begin{equation}
    X_{RAE}=R(X,Mes),
    \label{alg:1}
\end{equation}
where $X$ is the carrier, and Mes is the embed message. In this paper, considering the balance between capacity and efficiency, we employ the LSB method for data embedding.


\section{Methodology}
\subsection{Overview}
\begin{figure*}[t]
  \centering
  \includegraphics[width=1\linewidth]{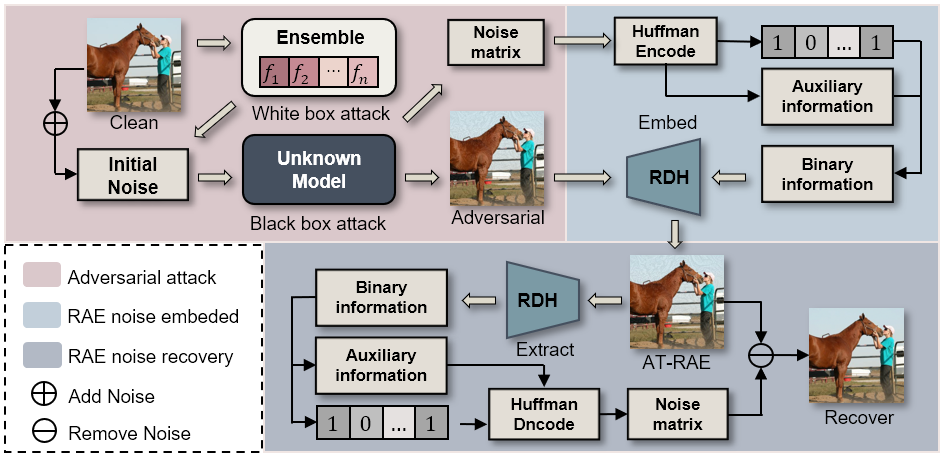}
  \caption{An overview of the proposed DP-TRAE.}
  \label{figure2} 
\end{figure*}
This section elaborates on the DP-TRAE framework, which comprises three core modules illustrated in Figure \ref{figure2}: the SA-WA module for white-box attacks, MAE-BA for black-box scenarios, and a reversible mechanism ensuring image preservation. DP-TRAE begins by performing rapid adversarial preprocessing on the input clean image to generate robust adversarial examples, thereby reducing the query overhead for the second-stage MAE-BA. MAE-BA estimates the gradient direction by querying superpixel blocks and utilizes historical query results to effectively improve attack efficiency. SA-WA and MAE-BA can be used independently to adapt to different attack scenarios. Finally, Huffman coding is applied to compress the information of the matrix, which is then embedded using RDH to generate DP-TRAE. During restoration, RDH extracts and reverses the perturbation matrix to losslessly recover the image.

\subsection{Preliminary}
Consider the white-box model $f$ and the unknown target model $b$, where $x \in \mathcal{X}$ is a benign input with dimensions $H \times W \times C$ and the corresponding ground-truth label $y \in \mathcal{Y}$. Let $f(x)$ and $b(x)$ denote the prediction results of the white-box and black-box models, respectively. Consistent with prior work, we assume that the complete gradient information of $f$ is available while $b$ is entirely unknown, providing only the output labels and the associated probabilities.

For white-box attacks, given a benign input $x$ and a loss function $\mathcal{J}$ (e.g., the Cross-Entropy loss), the aim of the attack can be formulated as:

\begin{equation}
    \delta = \arg \max \mathcal{J}(f(x+\delta), y) ,
    \label{alg:2}
\end{equation}

\begin{equation}
    f(x+\delta )\ne y ,\quad s.t.\left \| \delta \right \| _{\infty } < \epsilon, 
    \label{alg:3}
\end{equation}
where $\delta$ represents the adversarial perturbation and $\epsilon$ is a constant that controls the norm constraint. 

MI-FGSM is a classic attack algorithm that incorporates momentum to stabilize the optimization process to maximize Eq. \ref{alg:2} and improve attack success rates. In MI-FGSM, the adversarial perturbation is iteratively updated, and the accumulated gradient is used to determine the direction of the perturbation. Given a step size $\alpha$, a number of iterations $T$, and a decay factor $\mu$, the iterative update is defined as follows:

\begin{equation}
g_{t+1} = \mu \cdot g_t + \frac{\nabla_\delta \mathcal{J}(f(x_t + \delta_{t}), y)}{\left\| \nabla_\delta \mathcal{J}(f(x_t + \delta_{t}), y) \right\|_1},
\label{alg:4}
\end{equation}

\begin{equation}
\delta_{t+1} = \delta_t + \alpha \cdot \text{sign}(g_{t+1}),
\label{alg:5}
\end{equation}

where $g_t$ is the accumulated gradient at iteration $t$, and $\delta_t$ represents the adversarial perturbation at iteration $t$. The use of the momentum term $\mu$ helps in accumulating the gradient information from previous steps, thereby generating more effective perturbations. 

However, a single image input limits the robustness of the attack. Therefore, prior works have utilized a combination of DI and TI to increase input diversity, effectively preventing overfitting to a single model. And the iterative equation becomes:
\begin{equation}
g_{t+1}=\mu \cdot g_{t}+\frac{T\cdot \nabla _\delta \mathcal{J}\left ( f\left ( D\left ( x+\delta _{t} \right ),y  \right )  \right )  }{\left \| T\cdot \nabla _\delta \mathcal{J}\left ( f\left ( D\left ( x+\delta _{t} \right )  \right ),y  \right ) \right \|_{1} } ,
\label{alg:6}
\end{equation}
where $T$ is the convolution kernel in TI and $D$ is the diverse input transformation in DI. We employ the combination of DI, TI, and MI as the basis of the attack to gain more transferability.
\subsection{Stepwise Adaptive Attack}
The SA-WA module aims to produce highly transferable adversarial perturbations. In white-box attacks, internal details of the target model are accessible, which allows for the effective use of gradient information to craft adversarial perturbations. During this attack process, we observed that the gradients of the model exhibit non-uniform variations across different input regions. The SA-WA leverages this by applying more substantial perturbations to gradient-sensitive areas, thereby prioritizing regions where the model's decision boundaries are most susceptible to deviation. This approach enhances the efficiency of generating adversarial perturbations, allowing significant attack effects to be observed in the initial iterations.


However, focusing solely on enhancing perturbations in gradient-sensitive regions may lead to overfitting to a single model, resulting in suboptimal performance and reduced attack transferability. Therefore, we introduced the adaptive strategy for scaling the perturbations in sensitive regions. Specifically, as the number of iterations increases, we progressively reduce the amplification applied to these regions. By doing so, we mitigate the risk of overfitting during later iterations, ensuring a balance between attack effectiveness and model transferability, and the SA-WA algorithm is presented in Algorithm \ref{alg:simplified_dtmi_sa_attack}.

In addition to focusing on attack performance, it is essential to consider the embedding and recovery of perturbations in subsequent steps, as RDH technology imposes strict storage limitations, and the magnitude of adversarial perturbations at each position varies, which complicates direct encoding and storage. For instance, with a $\epsilon $ size of 8/255, storing the perturbation for a single pixel across three channels would require 48 bits, not even accounting for cases where the perturbation is zero. Therefore, it is crucial to compress the perturbation while preserving its effectiveness.

\begin{algorithm}[H]
\begin{spacing}{1}
\caption{DI-TI-MI-SA}
\label{alg:simplified_dtmi_sa_attack}
\begin{algorithmic}[1]
    \Require Clean image $x$, model $f$, loss function $\mathcal{J}$, Iterations $N$, step size $\alpha$, perturbation limit $\epsilon$, original label $y$
    \Ensure Adversarial perturbation $\delta$
    \State Initialize adversarial perturbation: $\delta = 0,X_{adv}=X$
    \For{$i \gets 0$ to $N-1$}
        \State $X_{adv}=X+\delta_{i}$
        \State Apply DI:$X_{adv}=D(X_{adv})$
        \State Calculate gradients: $g_{i+1}=\nabla_{\delta }f(x_{adv},y)$
        \State Apply TI,MI: utilize Eq. \ref{alg:6} and make $g_{temp}=g_{i+1}$
        \State Apply SA: The top $(N-i)/2N$ of $g_{temp}$ magnitude changes are labeled as 2, while the rest are labeled as 1.
        \State Update perturbation: $\delta_{i+1} = \delta_{i} + \alpha \cdot \text{sign}(g_{t+1})\cdot g_{temp}$
        \State Clip perturbation: $\delta_{i+1} = \text{clip}(\delta_{i+1}, -\epsilon, \epsilon)$
    \EndFor
    \State \Return $\delta$
\end{algorithmic}
\end{spacing}
\end{algorithm}

To address this, SA-WA incorporates a compression mechanism directly into the iterative attack process. By embedding the compression step, SA-WA minimizes the impact of compression on attack performance, allowing for efficient storage without significantly compromising the attack's effectiveness. SA-WA first aggregates the gradients of channels, reducing the three-dimensional perturbation $\delta$ to  one dimension:
\begin{equation}
   g_{t}=\mathbb{E}_{c \in \{R, G, B\}}[g_{c_{t}}] = \frac{1}{3} \sum_{c \in \{R, G, B\}} g_{c_{t}}.
   \label{alg:7}
\end{equation}

At this stage, the data storage volume is reduced to one-third of its original size. The perturbations are then compressed using a threshold-based approach, where precise values are replaced with uniform noise:

\begin{equation}
E(\left | \delta \right |  )=\begin{cases}
2,\frac{\epsilon}{2} \le \left | \delta \right | \le \epsilon
 \\1,0 <  \left | \delta \right | \le \frac{\epsilon}{2}
 \\0,\left | \delta \right |=0
\end{cases}
\label{alg:8}
\end{equation}

\begin{equation}
\delta _{t+1} = sign(\delta _{t+1})\cdot E(\left | \delta _{t+1}  \right | ) \cdot \xi ,
\label{alg:9}
\end{equation}
where $\xi $ represents the stage threshold used to compress the perturbation values within a piecewise distribution. Compressed perturbation $\delta$ will be applied as the initial perturbation matrix for the MAE-BA method.

\subsection{Memory-Assisted Expansion Attack}
For black-box models, the SimBA method for generating adversarial perturbations is a simple yet effective approach. SimBA iteratively applies perturbations in randomly selected directions and observes the resulting change in the target class probability. Perturbations that effectively reduce the target class probability are retained, enabling SimBA to reliably lower the model’s prediction confidence through repeated adjustments. This approach resembles finite difference methods for gradient estimation \cite{zoo}, but with a key distinction: SimBA's stochastic coordinate perturbations do not directly estimate gradients, instead leveraging observed outcomes to generate effective perturbations. In this work, we adopt stochastic coordinate perturbations as the default strategy.

However, pixel-level stochastic coordinate perturbations, which involve perturbing individual pixels, result in a high number of queries, leading to increased costs, especially when attacks are repeated over multiple iterations. To address this, the MAE-BA method introduces the concept of superpixel blocks to reduce query complexity and associated costs. Superpixel blocks group neighboring pixels into cohesive regions, allowing for coordinated exploration of the input image via larger pixel groups rather than individual units. Operating on these aggregated blocks significantly reduces the required queries while maintaining perturbation effectiveness.

\begin{algorithm}[H]
\begin{spacing}{1}
\caption{MAE-BA}
\label{alg:MAE-BA}
\begin{algorithmic}[1]
    \Require Clean image $x$, White adversarial noise $\delta$, Black model $b$, Iterations $N$, stage threshold $\xi$, perturbation limit $\epsilon$, original label $y$, enhance step size $s$, Expand size $Ep$
    \Ensure Adversarial perturbation $\delta$
    \State Initialize Empty Memory list $H$
   
    
        \For{$i \gets 0$ to $N-1$}
            \If{$b(x+\delta) \neq y$}
                \State Select random direction $q$
                \State Select random superpixel blocks $E_{j}$
                \State $P_{\text{pre}} = P_{b}(y|x+\delta)$
                \If{$(i+1)\%s=0$}
                    \State Replace index $j$ with the largest value in $H$
                    \State Expand superpixel blocks $E_{j}$ size with $Ep$
                    \State $H[j]=0$
                \EndIf
                \State $\delta = \delta + q \cdot \xi \cdot E_{j}$
                \State $P_{\text{next}} = P_{b}(y|x+\delta)$
                \If{$P_{\text{pre}} < P_{\text{next}}$}
                    \State $\delta = \delta - q \cdot \xi \cdot E_{j}$
                    \State $P_{\text{next}} = P_{b}(y|x+\delta)$
                \EndIf
                \State $H \gets (P_{\text{pre}}/P_{\text{next}},j)$
                
                \State Clip perturbation: $\delta = \text{clip}(\delta, -\epsilon, \epsilon)$
            \EndIf
        \EndFor
    \State \Return $\delta$
\end{algorithmic}
\end{spacing}
\end{algorithm}

In addition to employing superpixel blocks, we utilize information obtained from historical queries, a factor often neglected in typical black-box attack scenarios. Historical information provides insights into the sensitivity of specific image regions to perturbations, guiding the informed selection of subsequent query directions. Leveraging this historical knowledge enhances search efficiency by avoiding redundant queries and accelerating convergence toward effective adversarial perturbations. Specifically, we prioritize superpixel blocks previously identified as promising based on historical query data, expanding the exploration scope around these regions. By focusing on the neighborhoods of promising superpixel blocks, we exploit the intrinsic local coherence of the gradient, where small perturbations in adjacent regions tend to yield similar effects on model output. This approach reduces redundant queries in less sensitive areas, concentrating instead on regions where minor adjustments are more likely to induce significant changes in model behavior. Furthermore, integrating historical query information as adaptive feedback guides the attack towards areas of maximum vulnerability, improving both the effectiveness and query efficiency of the adversarial perturbation process and the whole algorithm is presented in Algorithm \ref{alg:MAE-BA}.

\subsection{Embed and Recover}
After compressing the perturbations using Eq. \ref{alg:7} and Eq. \ref{alg:8}, we observed a significant imbalance in the value distributions of different perturbations, as well as considerable variation among individual perturbations. Therefore, we decided to employ Huffman coding to effectively encode and store the perturbation information. By leveraging the advantages of Huffman coding, we can efficiently encode these perturbations with a higher compression ratio, thereby reducing storage requirements and improving processing efficiency. The detailed steps of the entire perturbation embedding algorithm are presented in Algorithm \ref{alg:embed}.

\begin{algorithm}[H]
\begin{spacing}{1}
\caption{Encode and Embed}
\label{alg:embed}
\begin{algorithmic}[1]
    \Require Clean image $x$, Adversarial noise $\delta$, stage threshold $\xi$, Flexible encryptor $F$, RDH technology $R$
    \Ensure Reversible Adversarial Examples DP-TRAE
    \State Initialize Huffman Tree $T$, Message matrix $Mse(H,W)$
    
    \For{$h \gets 0$ to $H-1$}
        \For{$w \gets 0$ to $W-1$}
            \State Calculation $stage$: $stage=\delta_{hw}/\xi$
            \State $Mse_{hw}=\xi$
            \State Contract Huffman Tree: $T.append(stage)$
        \EndFor
    \EndFor
    \State Compress 2D $Mes$ into 1D
    \State Encrypted message: $Mse=F(Mes+T)$
    \State Generate DP-TRAE: DP-TRAE=$R(x+\delta,Mse)$
    \State \Return DP-TRAE
\end{algorithmic}
\end{spacing}
\end{algorithm}

When performing the recovery operation, we only need to reverse the embedding process to losslessly restore the perturbation information. Furthermore, during the encoding, compression, and storage processes, we can flexibly introduce specific encryption measures to ensure that the information remains secure during transmission and storage. 
\section{Experiment}
\subsection{Experiment setup}
\textbf{Dataset and Environment.} In this study, we employed the ILSVRC2012 dataset \cite{russakovsky2015imagenet} to assess the effectiveness of various adversarial attack methods across different deep learning models. We randomly selected 1,000 images that the target models could classify accurately. All experiments were conducted on an NVIDIA A40 GPU, which ensured efficient processing capabilities for the extensive computations involved.

\begin{table*}[t]

\caption{ASR (\%) on several models under attack scenarios using Res-50, VGG-16, and Inc-v3 as the white-box models, respectively.}
\resizebox{\textwidth}{!}{%
\begin{tabular}{lccccccccccc}

\toprule[0.15em]
\textbf{Source : RN-50} & \multicolumn{10}{c}{Target model} &  \\
\cmidrule(l){2-11}
Attack & RN-34 & RN-50 & RN-152 & DN-121 & VGG-16 & VGG-19 & Inc-v3 & Alexnet & Mob-v2 & Mob-v3 & Average \\ \midrule
Liu \cite{RAEliu2023unauthorized}   & 27.8 & 99.9 & 29.0 & 26.1 & 26.4 & 23.8 & 11.4 & 11.7 & 21.2 & 6.6  & 28.4 \\
RIT \cite{RAEyin2019reversible}  & 34.7 & 99.8 & 31.3 & 31.8 & 26.8 & 27.0 & 15.1 & 19.1 & 24.0 & 9.4  & 31.9  \\
DP-RAE \cite{DP-RAE} & 47.1 & \textbf{100}  & 43.9 & 44.7 & 45.2 & 46.6 & 24.1 & 26.7 & 43.5 & 16.5 & 43.8 \\
DP-TRAE (Ours) & \textbf{71.3} & 99.3 & \textbf{68.4} & \textbf{74.3} & \textbf{81.0} & \textbf{81.4} & \textbf{50.3} & \textbf{37.6} & \textbf{73.7} & \textbf{35.1} & \textbf{67.2} \\

\toprule[0.1em]
\textbf{Source : Inc-v3} & \multicolumn{10}{c}{Target model} &  \\
\cmidrule(l){2-11}
Attack & RN-34 & RN-50 & RN-152 & DN-121 & VGG-16 & VGG-19 & Inc-v3 & Alexnet & Mob-v2 & Mob-v3 & Average \\ \midrule
Liu \cite{RAEliu2023unauthorized}    & 5.0  & 4.0  & 4.0  & 4.9  & 6.2  & 6.9  & 97.3 & 9.2  & 7.3  & 3.9  & 14.9 \\
RIT \cite{RAEyin2019reversible}  & 9.7  & 8.4  & 6.8  & 9.6  & 9.8  & 10.7 & 97.6 & 15.2 & 12.3 & 8.1  & 18.8 \\
DP-RAE \cite{DP-RAE}  & 20.1 & 18.5 & 14.1 & 18.2 & 19.2 & 19.2 & \textbf{98.1} & 25.4 & 25.3 & 15.4 & 27.4 \\
DP-TRAE (Ours) & \textbf{41.5} & \textbf{40.1} & \textbf{32.5} & \textbf{44.6} & \textbf{45.1} & \textbf{47.3} & 96.9 & \textbf{30.5} & \textbf{45.4} & \textbf{24.6} & \textbf{44.9} \\

\toprule[0.1em]
\textbf{Source : VGG16} & \multicolumn{10}{c}{Target model} &  \\
\cmidrule(l){2-11}
Attack & RN-34 & RN-50 & RN-152 & DN-121 & VGG-16 & VGG-19 & Inc-v3 & Alexnet & Mob-v2 & Mob-v3 & Average \\ \midrule
Liu \cite{RAEliu2023unauthorized}    & 10.2 & 9.1  & 5.7  & 11.6 & 99.5 & 59.6 & 6.9  & 10.4 & 16.4 & 5.5  & 23.5 \\
RIT \cite{RAEyin2019reversible}  & 11.8 & 11.1 & 5.6  & 14.5 & 99.6 & 53.5 & 9.5  & 15.2 & 16.2 & 7.5  & 24.5 \\
DP-RAE \cite{DP-RAE}  & 16.2 & 17.2 & 10.9 & 17.6 & 99.7 & 79.4 & 12.3 & 21.9 & 26.2 & 12.5 & 31.4 \\
DP-TRAE (Ours) & \textbf{30.7} & \textbf{31.7} & \textbf{21.3} & \textbf{35.5} & \textbf{100}  & \textbf{87.4} & \textbf{21.4} & \textbf{33.7} & \textbf{47.0} & \textbf{23.0} & \textbf{43.2} \\
            \bottomrule
\end{tabular}%
}
\label{tab:table1}
\end{table*}

Regarding model selection, we focused on a range of models with diverse architectures and their corresponding sub-models, including ResNet34 (RN-34) \cite{resnet}, ResNet50 (RN-50) \cite{resnet}, ResNet152 (RN-152) \cite{resnet}, DenseNet121 (DN-121) \cite{densenet121}, VGGNet16-BN (VGG16) \cite{vgg16}, VGGNet19-BN (VGG19) \cite{vgg16}, Inception-v3 (Inc-v3) \cite{inception3}, Alexnet \cite{AlexNet}, MobileNet-v2 (Mob-v2) \cite{mobilenetv2}, MobileNet-v3 (Mob-v3) \cite{mobilenetv3}. This approach allowed us to evaluate the transferability of adversarial attacks across different model types, providing insights into their robustness and vulnerabilities.

\textbf{Attack Setting.} 
We set the maximum perturbation $\epsilon$ for the attack to $8/255$, the stage threshold was set to half of the $\delta$, and the step size $\alpha$ equals the stage threshold. The number of iterations for the white-box attack was set to $10$, while the maximum number of iterations for the black-box attack was set to $1000$. For the black-box attack, the expand size $Ep$ was set to $4$, with an enhanced step size as $s=5$.
 
\textbf{Evaluation Metrics.}
Regarding the evaluation of attack performance, we employed Attack Success Rate (ASR) as the primary metric to assess the effectiveness of misleading different models. ASR is defined as the proportion of images that successfully deceive the target model out of the total number of input images. The higher ASR values reflect greater attack performace.

For recovery performance, we employed Peak Signal-to-Noise Ratio (PSNR) and Structural Similarity Index (SSIM) \cite{SSIM}, which are widely used metrics for evaluating image quality. The higher PSNR value implies that the restored image more closely resembles the original. On the other hand, SSIM assesses the similarity between two images and with values ranging from 0 to 1. A higher SSIM value indicates a more remarkable similarity to the original image regarding structural details. Together, these metrics provide a comprehensive assessment of the visual quality of the recovered images.

In addition, we evaluated the recognition accuracy of the recovered images, referred to as Success Rate (SR). SR measures the proportion of recovered images that the model correctly classifies. This metric provides insight into whether adversarial samples, after undergoing the recovery process, can successfully restore the classifier’s predictions to their original labels. A higher SR indicates that the recovery process effectively mitigates the impact of adversarial attacks, restoring the model's ability to correctly classify the images while preserving high visual quality.
\subsection{Attack performance}
In this section, we evaluate the attack performance of DP-TRAE from two perspectives: the attack performance of DP-TRAE in white-box scenarios and the attack performance of DP-TRAE in black-box scenarios.

\begin{table*}[h]
\caption{{The ASR (\%) on several models under attack scenarios using Res-50, VGG-16, and Inc-v3 as the ensemble models.}}
\footnotesize
\resizebox{\textwidth}{!}{%
\begin{tabular}{lccccccc|cccc}
\toprule[0.15em]
 & \multicolumn{7}{c}{Test model} & \multicolumn{3}{c}{Ensemble model} \\
\cmidrule(l){2-8}\cmidrule(lr){9-11}
Attack & RN-34 & RN-152 & DN-121 & Mob-v2 & Mob-v3 & VGG-19 & AlexNet & RN-50 & Inc-v3 & VGG-16 \\ \midrule
Liu \cite{RAEliu2023unauthorized}    & 29.5 & 28.8 & 34.8 & 31.0 & 6.5  & 59.5 & 9.5  & 99.1 & 99.0 & 99.4 \\
RIT \cite{RAEyin2019reversible}  & 43.4 & 36.8 & 42.9 & 36.8 & 14.1 & 63.8 & 20.4 & 99.1 & 98.8 & 99.4 \\
DP-RAE \cite{DP-RAE}  & 60.4 & 57.3 & 62.7 & 61.1 & 22.0 & 86.5 & 30.8 & \textbf{99.6} & \textbf{99.3} & 99.7 \\
DP-TRAE (Ours) & \textbf{83.2} & \textbf{81.4} & \textbf{87.6} & \textbf{85.6} & \textbf{44.4} & \textbf{96.2} & \textbf{42.9} & 99.2 & 99.0 & \textbf{99.9} \\

            \bottomrule
            
\end{tabular}%
}
\label{tab:table2}
\end{table*}

\begin{table*}[h]
\caption{The ASR (\%) on different models, with DN-121 as the black-box model, and DP-TRAE utilizing the ensemble preprocessing operation based on SA-WA.}
\resizebox{\textwidth}{!}{%
\begin{tabular}{lccccccccccc}

\toprule[0.15em]
 & \multicolumn{10}{c}{Target model} &  \\
\cmidrule(l){2-11}

Attack & RN-34 & RN-50 & RN-152 & DN-121 & VGG-16 & VGG-19 & Inc-v3 & Alexnet & Mob-v2 & Mob-v3 & Average \\ \midrule
Simba \cite{Simba}     & 1.0  & 0.4  & 0.2  & 94.6 & 1.2  & 1.0  & 0.9  & 1.7  & 0.3  & 0.2  & 10.2 \\
Simba-DCT \cite{Simba} & 0.7  & 0.6  & 0.5  & 95.9 & 0.9  & 0.8  & 1.5  & 2.6  & 0.7  & 1.0  & 10.5 \\
Surfree \cite{Surfree}   & 13.3 & 11.6 & 4.1  & 97.2 & 24.8 & 21.6 & 8.2  & 54   & 33.8 & 23.6 & 29.2 \\
DP-RAE \cite{DP-RAE}    & 60.2 & \textbf{99.5} & 56.4 & \textbf{99.0} & 99.5 & 82   & \textbf{99.1} & 29.6 & 60.2 & 22.0 & 70.8 \\
DP-TRAE (Ours)    & \textbf{80.9} & 98.0 & \textbf{79.7} & \textbf{99.0} & \textbf{99.6} & \textbf{92.2} & 97.1 & \textbf{41.9} & \textbf{83.8} & \textbf{43.2} & \textbf{81.5} \\
            \bottomrule
\end{tabular}%
}

\label{tab:table3}
\end{table*}

\textbf{White-box scenarios.} In this experimental setup, we compared DP-TRAE with several existing RAEs, including the RAE \cite{RAEliu2023unauthorized} proposed by Liu \textit{et al.}, the RAE based on Reversible Image Transformation (RIT) \cite{RAEyin2019reversible} and our previous work DP-RAE \cite{DP-RAE}.  We selected three structurally diverse models to conduct the attacks and used it to generate adversarial examples, which were then tested on the other models to assess single-model transferability. Table \ref{tab:table1} shows the results. First, the proposed DP-TRAE consistently outperformed across most scenarios. This is because that the method proposed by Liu \textit{et al.} fails to compress the perturbation magnitude, resulting in over-detailed storage processes that significantly increase the steganographic storage overhead. This issue becomes particularly pronounced when the $\epsilon$ parameter is large, forcing Liu's approach to sacrifice partial attack performance while meeting steganographic requirements. The RIT method achieves reversible adversarial attacks by directly disguising the original image as adversarial examples. However, its inability to losslessly preserve adversarial perturbation details during sample generation ultimately compromises the effectiveness of the attack. Our previous work DP-RAE employs grayscale-invariant steganography. While maintaining image grayscale properties, it significantly reduced storage efficiency and DP-RAE did not compress the perturbation information effectively. To preserve attack efficiency, the method relies on super-pixel blocks for perturbation compression, which inevitably degrades attack performance and leads to limited cross-model transferability. In contrast, DP-TRAE adopts a more concise and efficient RDH technique while leveraging Huffman coding to compress the perturbations. This approach eliminates the need for additional regional compression, effectively preserves attack performance, and demonstrates superior transferability compared to DP-RAE. Additionally, we observed that perturbations generated using the RN-50 exhibited better transferability across different models. It is attributed to the residual connections of RN-50, which are used to capture hierarchical and generalized feature representations. Residual connections help preserve crucial information across layers, resulting in perturbations that generalize better, making them more effective when used to deceive other models. Inc-V3 relies on convolutional blocks that focus on multi-scale feature extraction, which may result in more specialized perturbations to that specific architecture and be less effective on other models with different structures. Similarly, VGG16, with its simpler and more uniform convolutional stack, may lack the ability to generate perturbations that capture complex, transferable features, leading to a reduced effectiveness when transferred to dissimilar models. Moreover, it can be observed that the generated perturbations demonstrate stronger transferability when applied to homologous models. Through the above analysis, it is suggested that both the architectural characteristics of the model used to generate adversarial examples and the nature of the learned feature representations play a crucial role in determining the transferability of adversarial attacks. 

To further assess the transferability of adversarial perturbations, we employed an ensemble attack strategy, which enhances the cross-model transferability by optimizing perturbations across multiple models simultaneously. The perturbations were generated by integrating several white-box models, each assigned with simple weighting factors. As presented in Table \ref{tab:table2}, the results demonstrate that the DP-TRAE consistently improves ASR in all cases. This improvement is attributed to introducing the stage threshold and Huffman coding compression mechanism, which effectively reduces the storage requirement for unit perturbations while ensuring the success rate and effectiveness of the attack, leading to a more refined application of perturbations. Additionally, the extra perturbation in gradient-sensitive regions accelerates the generation process.

\textbf{Black-box attack scenario.} Due to the current lack of research on black-box reversible attack methods, we compared DP-TRAE with several existing state-of-the-art black-box attack methods, including Simba \cite{Simba}, Simba-DCT \cite{Simba}, Surfree \cite{Surfree}, and our previous work DP-RAE \cite{DP-RAE}. The experimental results, as shown in Table \ref{tab:table3}, indicate that although these query-based black-box attacks exhibit strong performance when targeting specific models, they generally suffer from poor transferability. Specifically, these methods typically rely on optimizations tailored to specific models, resulting in suboptimal performance when transferred to different target models. The main reason for the poor transferability is that these methods fail to effectively leverage the common features across different models, leading to a significant degradation in attack performance during cross-model transfer. In contrast, DP-TRAE enhances transferability by introducing preprocessing noise generated during the white-box phase. The introduction of this noise provides a more stable perturbation pattern for subsequent attacks, allowing DP-TRAE to achieve more consistent attack performance across different models. In addition, the MA-EA method further enhances the attack effect on the target model by optimizing the perturbation through the historical query results. Compared to the previous version DP-RAE, DP-TRAE produces black-box attack with better transferability due to the improved attack performance of the white-box attack phase.

\begin{figure*}[t]
  \centering
  \includegraphics[width=0.9\linewidth]{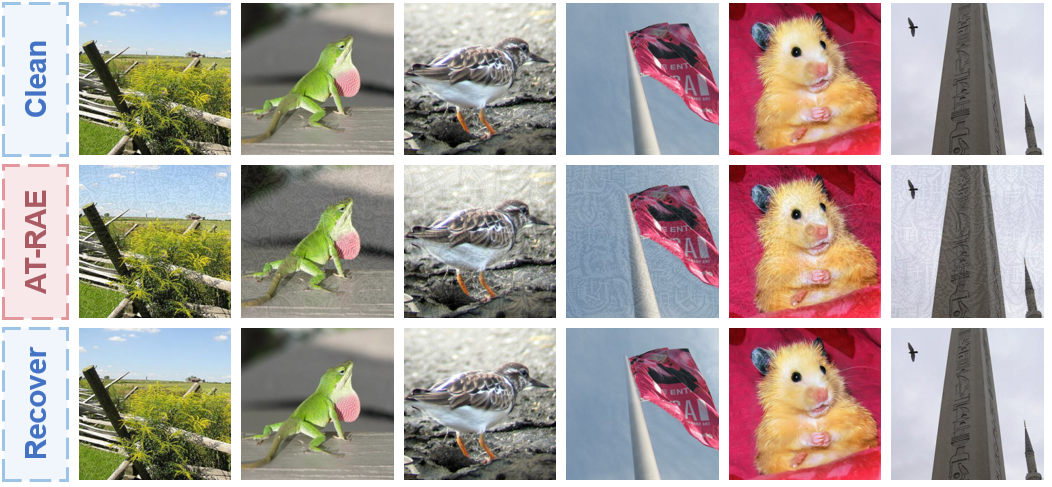}
  \caption{Visual results of DP-TRAE, including clean images, attacked images, and the corresponding recovery images.}
  \label{figure3} 
\end{figure*}

\subsection{Robustness evaluation}
In practical applications, networks often preprocess input data and apply defense techniques to reduce the impact of adversarial attacks, improving overall performance. These defenses aim to reduce the effectiveness of adversarial examples, ensuring the model remains stable and accurate under malicious perturbations. Therefore, the robustness of adversarial examples against these defenses is crucial.
In this study, we employ several preprocessing and defense methods, including Spatial Squeezing (Spatial) \cite{jpeg}, Random Resizing and Padding (Random) \cite{Random}, Gaussian Blurring (Gaussian) \cite{gaussian}, JPEG Compression (JPEG) \cite{jpeg}, and Super-resolution (Super) \cite{Super}.

\begin{table}[h]
\caption{{The ASR (\%) of adversarial attacks when againsting different defence methods.}}
\label{tab:table4}
\resizebox{\linewidth}{!}{
\centering
\begin{tabular}{lccccc}
\toprule[0.15em]
\multirow{2}{*}{\begin{tabular}[c]{@{}c@{}}Attack\\ method\end{tabular}} & \multicolumn{5}{c}{Defense method} \\
\cmidrule(l){2-6}
& Spatial & Random & Gaussian & JPEG & Super \\
\midrule

Simba \cite{Simba}     & 4.5  & 16.8 & 20.3 & 20.3 & 2.7  \\
Simba-DCT \cite{Simba} & 7.8  & 32.6 & 32.4 & 23.5 & 8.0  \\
Surfree \cite{Surfree}   & 20.0 & 23.4 & 23.7 & 54.0 & 13.6 \\
DP-RAE \cite{DP-RAE}    & 63.1 & 52.0 & 39.6 & 52.3 & 74.0 \\
DP-TRAE (Ours)    & \textbf{89.9} & \textbf{54.5} & \textbf{50.7} & \textbf{82.3} & \textbf{88.3} \\

\bottomrule
\end{tabular}}
\end{table}

As shown in Table \ref{tab:table4}, adversarial perturbations from black-box attacks lose their effectiveness when subjected to these defensive techniques. This is because such preprocessing introduces uncertainty, weakening the impact of query-based attacks.
In contrast, DP-TRAE utilized the white-box preprocessing approach to identify and exploit shared vulnerabilities across different models, thereby reducing the effectiveness of defense strategies. This approach enhances robustness by addressing vulnerabilities common to multiple model architectures.

\subsection{Recover ability}
RAEs are often evaluated based on their recovery performance. Notably, previous studies overlooked the assessment of this key metric. To demonstrate the recovery capability of DP-TRAE, we compared the images restored by DP-TRAE with the clean samples. As shown in Figure \ref{figure3}, the perturbations introduced by DP-TRAE cause only a slight degradation in image quality, which is imperceptible to human observers but can be devastating to adversarial models. The restored images effectively eliminate these perturbations without loss, making them indistinguishable from the original images.

\begin{table}[h]
\centering
\caption{{We reported the PSNR and SSIM of the DP-TAE, DP-TRAE and its recoverd examples, and we also evaluated their classfication success rate (SR) in the target model, "$\uparrow$" means the bigger the better.}}
\small
\label{tab:table5}
\resizebox{0.8\linewidth}{!}{
\begin{tabular}{lccc}
\toprule[0.15em]

& PSNR (dB) $\uparrow$ & SSIM $\uparrow$ & SR (\%) $\uparrow$ \\
\midrule 
DP-TAE & 31.14 & 0.8256 & 1.1   \\
DP-TRAE & 31.10 & 0.8242 & 1.1   \\
Recover (DP-TRAE) & \textbf{48.94} & \textbf{0.9913} & \textbf{100}   \\
\bottomrule
\end{tabular}}
\end{table}

Table \ref{tab:table5} presents an evaluation of the restored images using several quality metrics. The PSNR of the restored images exceeds 45 dB, and the SSIM approaches 1 and the recovered images successfully recover the correct classifications in the model. In addition, we observed that the DP-TAE and the recoverable DP-TRAE are almost identical in all respects, this is because of the threshold-based perturbation compressing and Huffman coding compression, the RDH approach does not have a large impact on the adversaral examples. These results demonstrate the effectiveness of DP-TRAE is not only maintaining the adversarial performance but also ensuring that the restored images retain both high visual fidelity and model accuracy.

\begin{table*}[t]
\renewcommand{\arraystretch}{1.2}
\caption{{ASR (\%) on several models under attack scenarios using Res-50, VGG-16, and Inc-v3 as the ensemble models. we conduct these experiments under two methods and report the ASR with Original/SA-WA method.}}
\footnotesize
\resizebox{\textwidth}{!}{%
\begin{tabular}{lccccccc|cccc}
\toprule[0.15em]
\textbf{steps = 5} & \multicolumn{7}{c}{Test model} & \multicolumn{3}{c}{Ensemble model} \\
\cmidrule(l){2-8}\cmidrule(lr){9-11}
Attack & RN-34 & RN-152 & DN-121 & VGG-19 & AlexNet  & Mob-v2 & Mob-v3& RN-50 & VGG16 & Inc-v3 \\ \midrule

BIM     & 45.0/\textbf{48.3} & 40.7/\textbf{42.5} & \textbf{47.8}/47.2 & 73.4/\textbf{75.9} & 18.9/\textbf{21.9} & 44.7/\textbf{48.9} & 14.7/\textbf{16.1} & \textbf{98.6}/\textbf{98.6} & 99.3/99.3 & 98.8/98.6 \\
MI-FGSM & 56.7/\textbf{58.3} & 51.1/\textbf{52.2} & 58.6/\textbf{58.9} & 82.1/\textbf{82.6} & 27.7/\textbf{30.8} & 57.2/\textbf{57.8} & 21.0/\textbf{21.9} & \textbf{99.3}/99.2 & \textbf{99.4}/99.3 & 98.6/\textbf{98.8} \\
DI-MI   & 60.2/\textbf{62.1} & 55.2/\textbf{57.0} & 64.1/\textbf{65.4} & 86.0/\textbf{87.9} & 33.0/\textbf{34.9} & 64.0/\textbf{67.2} & 25.8/\textbf{26.7} & 90.1/\textbf{92.1} & 98.4/\textbf{99.7} & 88.7/\textbf{91.5} \\
DTMI    & 67.9/\textbf{70.3} & 63.5/\textbf{64.8} & 71.9/\textbf{75.0} & \textbf{89.7}/89.4 & 39.9/\textbf{40.3} & 71.4/\textbf{75.6} & 38.6/\textbf{40.6} & 91.1/\textbf{92.9} & \textbf{99.2}/98.0 & 90.8/\textbf{94.1} \\

\midrule[0.1em]
\textbf{steps = 10} & \multicolumn{7}{c}{Test model} & \multicolumn{3}{c}{Ensemble model} \\
\cmidrule(l){2-8}\cmidrule(lr){9-11}
Attack & RN-34 & RN-152 & DN-121 & VGG-19 & AlexNet  & Mob-v2 & Mob-v3& RN-50 & VGG16 & Inc-v3 \\ \midrule

BIM     & 53.2/\textbf{54.9} & 47.3/\textbf{52.2} & 51.8/\textbf{56.5} & 81.7/\textbf{82.8} & 21.4/\textbf{21.7} & 53.1/\textbf{54.6} & 16.3/\textbf{17.9} & \textbf{99.5}/\textbf{99.5} & \textbf{99.6}/99.4 & 99.0/\textbf{99.2} \\
MI-FGSM & 58.5/\textbf{60.9} & 55.0/\textbf{56.6} & 61.2/\textbf{63.4} & 85.2/\textbf{86.2} & 29.4/\textbf{30.1} & 59.2/\textbf{61.3} & 21.5/\textbf{23.1} & 99.5/\textbf{99.6} & 99.5/\textbf{99.6} & \textbf{99.3}/\textbf{99.3} \\
DI-MI   & 72.8/\textbf{76.6} & 69.7/\textbf{74.5} & 78.0/\textbf{80.4} & 93.9/\textbf{95.0} & 36.5/\textbf{36.7} & 77.3/\textbf{79.2} & 29.9/\textbf{30.9} & 96.8/\textbf{98.6} & \textbf{99.8}/\textbf{99.8} & 96.2/\textbf{97.9} \\
DTMI    & 80.5/\textbf{84.9} & 78.3/\textbf{82.8} & 83.9/\textbf{86.6} & 95.1/\textbf{96.2} & 43.1/\textbf{43.4} & 84.7/\textbf{87.9} & 43.6/\textbf{44.7} & 98.3/\textbf{99.3} & 99.8/\textbf{100}  & \textbf{98.7}/98.6 \\

            \bottomrule
            
\end{tabular}%
}

\label{tab:table6}

\end{table*}

\begin{figure*}[t]
  \centering
  \includegraphics[width=1\linewidth]{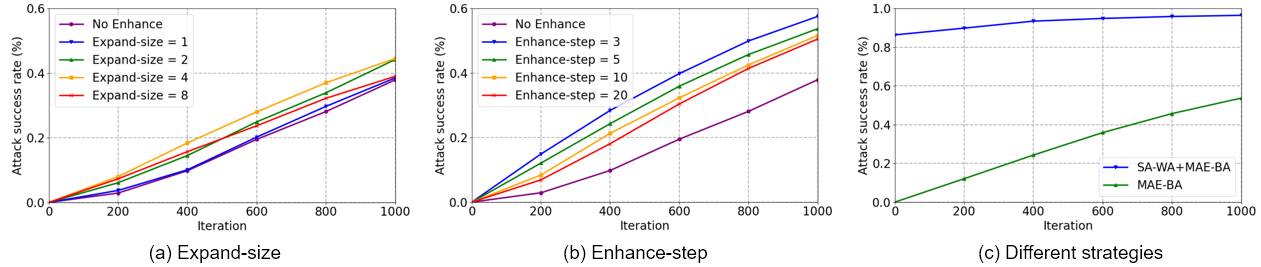}
  \caption{(a) reported the impact of different expand sizes on attack performance; (b) reported the impact of different enhance step sizes on attack performance; (c) reported the impact of dual-phase strategie on attack performance. }
  \label{figure4} 
\end{figure*}

\subsection{Ablation study}
This section presents an ablation study on DP-TRAE to evaluate the impact of different parameters and strategies on attack performance. 

\textbf{SA-WA:} First, we assessed the impact of the SA-WA on attack performance. To this end, we employed various gradient calculation strategies, including the IFGSM, MI-FGSM, DI-MI-FGSM (DI-MI), and DI-TI-MI-FGSM (DTMI). The core idea of these methods is to improve the efficiency of gradient calculation and the effectiveness of perturbations through different strategies, thereby enhancing the attack performance of adversarial examples.

As shown in Table \ref{tab:table6}, the SA-WA consistently demonstrated significant improvements in attack performance across different iteration counts. This indicates that SA, by applying additional perturbations to gradient-sensitive regions, can more effectively exploit model vulnerabilities, thereby increasing attack success rates and robustness, while accelerating the generation process of adversarial examples. Moreover, input diversity also played a critical role in gradient calculation, particularly in the DI-MI and DTMI methods. The input diversity effectively prevented gradients from falling into local optima, making the computed gradients more general and effective. This input diversification strategy enables adversarial examples to maintain a high success rate when facing different target models and defense mechanisms.

Overall, the SA-WA can be flexibly integrated with most gradient-based update methods, demonstrating its versatility and adaptability. This flexibility allows DP-TRAE to seamlessly incorporate different gradient calculation techniques as they evolve, enabling it to adopt more effective strategies to further improve attack performance. 

\textbf{MAE-BA:} Subsequently, we conducted an ablation study on the MAE-BA method to investigate the impact of varying enhancement frequency and expansion size on attack performance. Specifically, we examined how different settings for expanding perturbation regions and increasing enhancement frequency affected the adversarial attack efficacy. Notably, it can be seen as our previous work DP-RAE when the expand size is $0$. For clarity, we only utilized clean perturbations without employing white-box preprocessing techniques. The experimental results, as illustrated in Figure Figure \ref{figure4} (a), indicate a significant improvement in attack performance with an increase in the expansion size. 

The underlying mechanism can be attributed to the targeted amplification of perturbations around the points that historically demonstrated the highest effectiveness. By selectively enhancing perturbations in these key regions, the attack gains a more precise impact, thereby effectively leveraging the model's vulnerabilities. However, when the expansion size was allowed to grow without restriction, a decline in attack performance was observed. This deterioration can be explained by the fact that an overly extensive expansion range tends to blur the accurate gradient update direction, leading to the inclusion of numerous irrelevant points and resulting in incorrect gradient updates. Consequently, the efficacy of the attack DI-MInishes as the focus on key perturbation areas becomes diluted.

Moreover, we also observed a notable increase in attack success rate as the historical enhancement frequency was increased in Figure \ref{figure4} (b). This suggests that frequent reinforcement of previously effective perturbations allows for more persistent and accumulative exploitation of model weaknesses. However, considering the computational overhead associated with frequent updates, we opted to update the perturbations every five iterations. This approach strikes a balance between maintaining a high attack success rate and minimizing the additional computational burden. By selectively tuning both the expansion size and enhancement frequency, the MAE-BA method demonstrated an effective trade-off, achieving robust attack results while managing computational efficiency.

Finally, we tested the impact of different strategy combinations on the attack performance. As shown in Figure \ref{figure4} (c), when the SA-WA is combined with MAE, the attack efficiency improves significantly. This demonstrates that using adversarial perturbations as the initial disturbance can notably enhance performance in black-box model attacks. The combination of these strategies not only improves the precision of the attack but also allows for more effective exploitation of model vulnerabilities, leading to a higher success rate in bypassing the defenses of the black-box models. This result highlights the potential of adversarial perturbations in strengthening the performance of attacks under challenging black-box scenarios.

\subsection{Commercial model attack}
To evaluate the effectiveness of our RAE on real-world systems, we conducted tests on Baidu's cloud vision API\footnote[1]{\url{https://ai.baidu.com/tech/imagerecognition/general}}, an object recognition service. The objective of the attack was to mislead the top-3 categories returned by the API, all while adhering to the constraints of limited queries and perturbations. We selected 50 images for testing and achieved a 92\% success rate. Notably, a significant portion of the images were misclassified even before the queries were completed. This underscores the efficiency of the attack strategy, as the white-box perturbations applied at the outset were sufficiently powerful to influence the model's decision boundaries early in the querying process. Such early-stage perturbations highlight the potential effectiveness of adversarial attacks, particularly in scenarios with constrained query budgets.

\begin{figure}[h]
  \centering
  \includegraphics[width=1\linewidth]{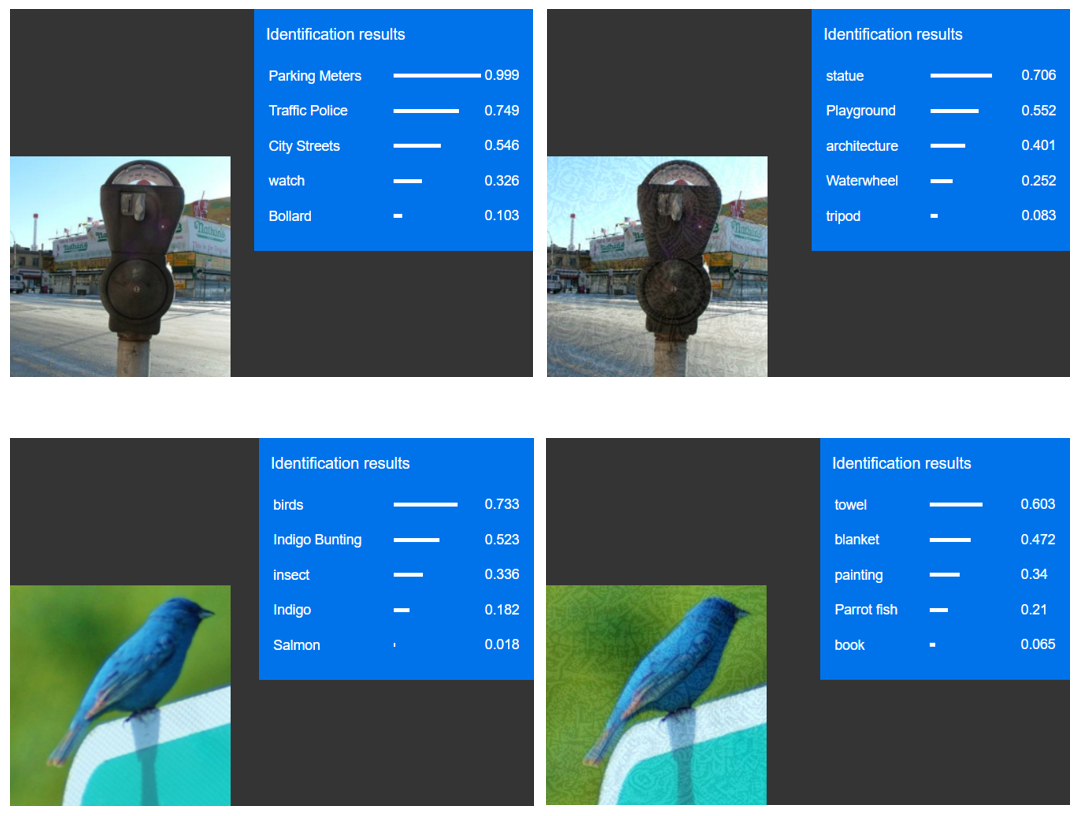}
  \caption{Results of DP-TRAE attacks on a commercial model.}
  \label{figure5} 
\end{figure}

As shown in Figure \ref{figure5}, DP-TRAE successfully misclassified the original labels, highlighting the potential threat of our method to commercial black-box models. Considering the limited number of queries allowed by commercial black-box models, we believe that increasing the number of queries can effectively enhance the success rate of the attack.

\section{Conclusion}
In this paper, we introduced the DP-TRAE method, which effectively combines the characteristics of different types of attacks to enhance the protection of sensitive data in complex environments. By leveraging the perturbations generated through white-box attacks, DP-TRAE significantly improves the transferability of adversarial examples, while the black-box attack component ensures targeted attacks on unknown models. Experimental results further demonstrate the superiority of our method, maintaining a high attack success rate even under various defense strategies. Notably, to the best of our knowledge, DP-TRAE is the first method to successfully perform reversible adversarial attacks on commercial black-box models. As a potential future direction, we are looking forward to extending our method to improve the performance of various applications such as large language models~\cite{hu2024agentscomerge,lin2025hsplitlora,fang2024automated} and distributed learning system~\cite{hu2024accelerating,zhang2024fedac,lin2025leo,zhang2024satfed}.

\section*{Acknowledgments}
This work was supported in part by the Xiamen Research Project for the Natural Science Foundation of Xiamen,China (3502Z202472028), the Xiamen Science and Technology Plan Project (3502Z20231042), the Xiamen University of Technology High-Level Talent Launch Project (YKJ22041R) and the Fundamental Research Funds for the Central Universities(1082204112364).

\bibliographystyle{unsrt}
\bibliography{hyref}

\end{document}